\documentclass[sigconf,table,prologue]{acmart}

\def\BibTeX{{\rm B\kern-.05em{\sc i\kern-.025em b}\kern-.08emT\kern-.1667em\lower.7ex\hbox{E}\kern-.125emX}}
    
\copyrightyear{2019}
\acmYear{2019}
\acmConference[PEARC'19]{PEARC '19: Practice and Experience in Advanced Research Computing}{Jul 28-- Aug. 1, 2019}{Chicago, IL}
\acmBooktitle{PEARC '19: Practice and Experience in Advanced Research Computing, Jul 28-- Aug. 1, 2019, Chicago, IL}

\begin{document}

\copyrightyear{2019} 
\acmYear{2019} 
\setcopyright{acmlicensed}
\acmConference[PEARC '19]{Practice and Experience in Advanced Research Computing}{July 28-August 1, 2019}{Chicago, IL, USA}
\acmBooktitle{Practice and Experience in Advanced Research Computing (PEARC '19), July 28-August 1, 2019, Chicago, IL, USA}
\acmPrice{15.00}
\acmDOI{10.1145/3332186.3332212}
\acmISBN{978-1-4503-7227-5/19/07}

%
\title{StashCache: A Distributed Caching Federation for the Open Science Grid}

%
\author{Derek Weitzel}
\email{dweitzel@unl.edu}
\orcid{0000-0002-8115-7573}
\author{Marian Zvada}
\email{zvada@unl.edu}
\affiliation{%
  \institution{University of Nebraska -- Lincoln}
  \streetaddress{1100 T St.}
  \city{Lincoln}
  \state{Nebraska}
  \postcode{68588}
}

\author{Ilija Vukotic}
\email{ivukotic@uchicago.edu}
\author{Rob Gardner}
\email{rwg@uchicago.edu}
\affiliation{%
  \institution{University of Chicago}
  \city{Chicago}
  \state{Illinois}}

\author{Brian Bockelman}
\email{bbockelman@morgridge.org}
\affiliation{%
  \institution{Morgridge Institute for Research}
  \city{Madison}
  \state{Wisconsin}
}

\author{Mats Rynge}
\email{rynge@isi.edu}
\affiliation{%
 \institution{Information Sciences Institute}
 \city{Los Angles}
 \state{California}}
 
\author{Edgar Fajardo Hernandez}
\email{emfajard@ucsd.edu}
\affiliation{%
 \institution{San Diego Super Computer Center}
 \city{La Jolla}
 \state{California}}
 
\author{Brian Lin}
\email{blin@cs.wisc.edu}
\author{M\'{a}ty\'{a}s Selmeci}
\email{matyas@cs.wisc.edu}
\affiliation{%
  \institution{University of Wisconsin -- Madison}
  \city{Madison}
  \state{Wisconsin}}

%
\renewcommand{\shortauthors}{Weitzel, et al.}

%
\begin{abstract}
Data distribution for opportunistic users is challenging as they neither own the computing resources they are using or any nearby storage.  Users are motivated to use opportunistic computing to expand their data processing capacity, but they require storage and fast networking to distribute data to that processing.  Since it requires significant management overhead, it is rare for resource providers to allow opportunistic access to storage.  Additionally, in order to use opportunistic storage at several distributed sites, users assume the responsibility to maintain their data.

In this paper we present StashCache, a distributed caching federation that enables opportunistic users to utilize nearby opportunistic storage.  StashCache is comprised of four components: data origins, redirectors, caches, and clients.  StashCache has been deployed in the Open Science Grid for several years and has been used by many projects.  Caches are deployed in geographically distributed locations across the U.S. and Europe.  We will present the architecture of StashCache, as well as utilization information of the infrastructure.  We will also present performance analysis comparing distributed HTTP Proxies vs StashCache.

\end{abstract}

%
\maketitle

\section{Introduction}

The Open Science Grid (OSG) provides access to computing resources distributed throughout the U.S. to perform data analysis.  A portion of users on the OSG are opportunistic users: they do not own the computing and storage resources which they use. While they continue to use the computing resources of the OSG, the data sizes have increased faster than the existing infrastructure can support.

Non-opportunistic users of the OSG are members of large experiments such as the Large Hadron Collider's ATLAS \cite{aad2008atlas} and CMS \cite{chatrchyan2008cms}.  These experiments own and maintain resources accessed through the OSG.  Each experiment manages the distribution of data to dozens of storage sites near the computing resources.  Since the experiments own storage, it is clear to users where to store their data and how to access data from the experiment.  Further, the large experiments have written complex software systems to manage the distributed storage.

On the other hand, opportunistic users do not have dedicated resources for their storage needs and therefore must use storage resources they do not own.  There are competing interests in the use of opportunistic resources, the resource owners and the opportunistic users.  The resource owner may want to reclaim space from the opportunistic user.  If the owner reclaims the space, the opportunistic user must discover that their data has been removed.  Further, the opportunistic user is responsible to transfer the data to multiple sites and to periodically check if the data is still available at these sites.  The opportunistic user has an interest to acquire the required storage for their workflow while not over-burdening the storage owner.

When a user's processing requires significant input data, they can transfer the data from the submit host or from a centralized server.  Transferring the data with each job puts network and I/O load on the submit node, starving it of resources to serve it's primary purpose to maintain the workflows run by the user. As the user increases the number of jobs running and requesting data, the submit node becomes the bottleneck.  Transferring data from a central service has similar bottlenecks as transferring from the submit host, though it leaves the submit host to only manage the user workflow.

Users of the OSG may use reverse HTTP proxies maintained by the large experiments such as CMS \cite{blumenfeld2008cms}.  The HTTP proxies are distributed across the OSG near the computing resources.  But, the HTTP proxies have well known limitations \cite{garzoglio2012supporting}.  One such limitation is that the proxies are optimized for small files such as software \cite{bockelman2014oasis} and experiment conditions \cite{dykstra2010greatly} rather than the multi-gigabyte files that some users require.  In this paper, we will present performance metrics for the use of StashCache over HTTP proxies for large files.

StashCache provides an easy interface for opportunistic users to distribute data to their processing.  It provides a caching layer between the researcher data location and the execution hosts requiring the data.  The caches are distributed throughout the OSG, and are guaranteed to have at least 10Gbps networking and several TB's of caching storage.




A cache provides many advantages over persistent storage space.  For users, a cache is simple to maintain since the data stored is transient and does not require maintenance.  The resource provider can reclaim space in the cache without worry of causing workflow failures when processing next tries to access the data.

\section{Background}


StashCache is comparable to a Content Distribution Network (CDN), which is used by hosting services to provide fast delivery of data to clients.  A CDN consists of a collection of (non-origin) servers that attempt to offload work from origin servers by delivering content on their behalf\cite{krishnamurthy2001use}. Popular commercial CDNs are Akamai\cite{nygren2010akamai}, Cloudflare\cite{cloudflare} and Fastly\cite{fastly}.

Content Distribution Networks have many of the same goals as StashCache:
\begin{itemize}
    \item Cache data from a set of origins
    \item Provide fast delivery of data to clients
    \item Cache near the client
\end{itemize}

Most CDNs are optimized for the delivery of web assets such as web pages, images and videos.  These items are usually small in size, less than a gigabyte.  For example, Cloudflare will only cache files up to 512MB \cite{cloudflare-cdn-largefile} and the cloud CDN Azure - Akamai service will by default only cache files up to 1.8GB \cite{azure-cdn-largefile}.  For some users of StashCache, the average file size is over 2GB.  Further, commercial CDNs do not have access to the high speed research networks within the U.S.


Accessing data available from many sources has been implemented in large experiments with frameworks such as Any data, any time, anywhere\cite{bloom2015any} and Federating ATLAS storage using XRootD \cite{gardner2014data}.  Federated storage allows clients to access data from several sources in a consistent namespace.  A client requests a file's location from a data location service, which will query storage servers to determine where the file is located.

\section{Implementation}

\begin{figure}
    \centering
    \includegraphics[width=\linewidth]{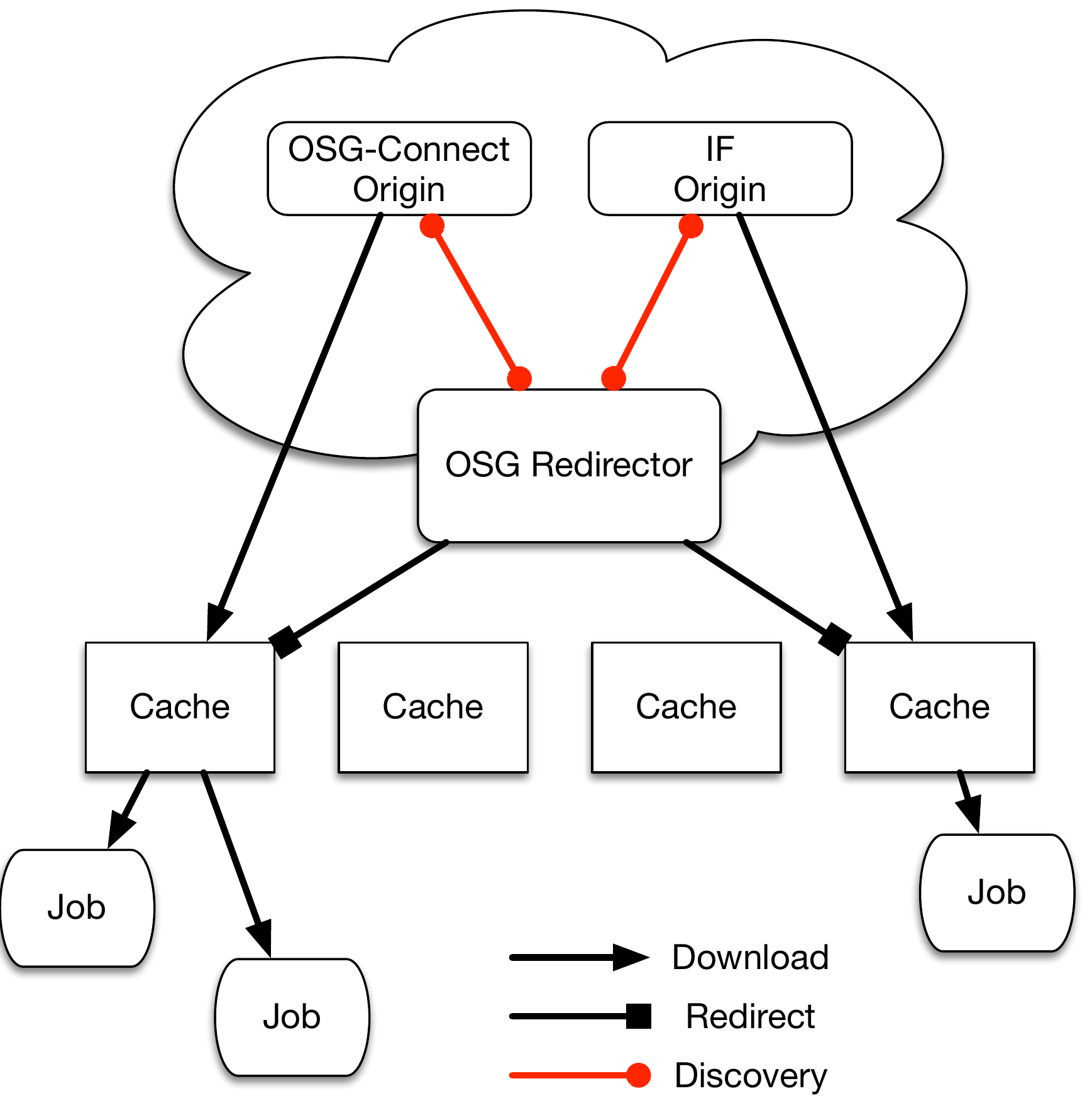}
    \caption{StashCache Architecture: Jobs request data from caches, which in turn query the redirector for the data location.  Data is transferred from the origin to the cache, and then to the job.}
    \label{fig:stashcache-architecture}
\end{figure}

The StashCache infrastructure consists of four components shown in Figure \ref{fig:stashcache-architecture}:
\begin{itemize}
    \item \textbf{Data Origin:} The authoritative source of data within the StashCache federation.  Each organization or experiment could have an origin.
    \item \textbf{Data Caches:} Regional servers that cache data from the origins before sending to the clients.
    \item \textbf{Redirector:} Serves as the data discovery service.  Caches query the redirector for the location of data.
    \item \textbf{Clients:} Requests data from the data caches.  The clients are responsible for finding the nearest cache using GeoIP.
\end{itemize}

Data origins are installed on the researcher's storage.  The origin is the authoritative source of data within the federation.  Each Origin is registered to serve a subset of the global namespace. Caches will contact the origin in order to retrieve and cache the data when it has been requested by a client.  The origin is a service, built upon the XRootD \cite{dorigo2005xrootd} software that exports datasets to the caching infrastructure.  The origin subscribes to a central XRootD redirector that is used by the caches to discover which origin has the requested data.

\begin{figure}[h]
	\centering
	\includegraphics[width=\linewidth]{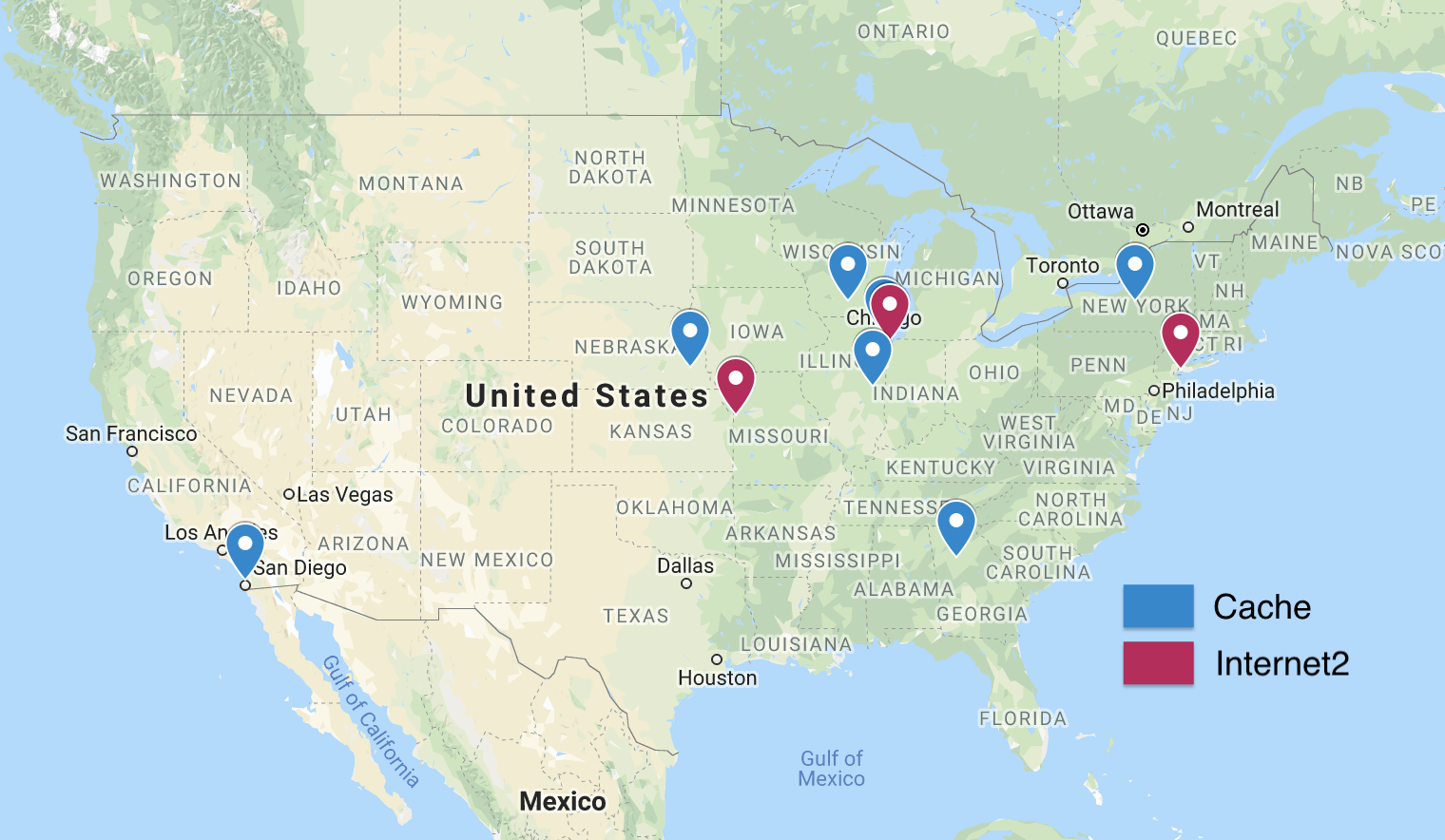}
	\caption{StashCache Locations within the U.S.  They are located at six universities and 3 Internet2 PoPs.}
	\label{fig:cachelocations}
\end{figure}

Caches also use XRootD to capture data requests from clients, download data from the origins, and to manage the cache space.  The caches receive data requests from the client, check the local cache, and if necessary locate and download the requested data from the origins.  The caches locate the data by contacting the XRootD redirector, which in turn asks the origins if they have the requested data.  Figure \ref{fig:cachelocations} shows the locations of the caches in the U.S.

The redirector serves as the data discovery service.  Caches query the redirector to find which origin contains the requested data.  The redirector will query the origins in order to find the data and return the the hostname of the origin that contains the data to the caches.  The redirector is a central service within StashCache hosted by the Open Science Grid.  There are two redirectors in a round robin, high availability configuration.

\subsection{Clients}
Two clients are used to read from the StashCache federation.  The CERN Virtual Machine File System (CVMFS) \cite{buncic2010cernvm} and stashcp \cite{derek_weitzel_2017_557111}.  In a previous paper, we highlighted how the Laser Interferometer Gravitational-Wave Observatory (LIGO) used the CVMFS client to access the StashCache federation \cite{weitzel2017data}.  

CVMFS provides a read-only POSIX interface to the StashCache federation.  It appears to users as a regular filesystem on the execute hosts.  When a user reads data from CVMFS, it will request the data from the caches and cache the data locally.  CVMFS determines which cache to contact by using its built-in GeoIP locator.

The posix interface that CVMFS provides allows it to view exactly what data the application requires.  CVMFS will download the data in small chunks of 24MB.  If an application only reads portions of a file, CVMFS will only download those portions.  

CVMFS is configured to only cache 1GB on the local hard drive.  There are two reason for this low cache amount.  The working set size of data transferred through the CVMFS client is expected to be too large to be reasonably accommodated on an execute node.  Also it is expected that transferring from a nearby cache is fast enough to serve the data to applications.

In order to provide a posix interface to remote storage, metadata such as the directory structure and size of files must be determined.  We wrote an indexer which will scan a remote data origin and gather metadata about the files.  Metadata includes:
\begin{itemize}
    \item File name and directory structure
    \item File size and permissions
    \item Checksum of files along the chunk boundaries
\end{itemize}

The indexer will detect changes to files by checking the file modification time and file size.  If it differs from the previously indexed file, it will reindex the file.  The indexer must scan the entire filesystem each iteration, causing a delay proportional to the number of files in the filesystem.

The second client, \texttt{stashcp}, is useful when CVMFS is not installed on the execute host or the delay caused by the indexing procedure is unacceptable for the user.  It is a simple script that will download data from the nearest cache, utilizing the CVMFS GeoIP infrastructure to find the nearest cache.  The interface to \texttt{stashcp} is similar to the Linux command \texttt{cp}.  

\texttt{stashcp} attempts 3 different methods to download the data:

\begin{enumerate}
    \item If CVMFS is available on the resource, copy the data from CVMFS
    \item If an XRootD client is available, it will download using XRootD clients.
    \item If the above two methods fail, it will attempt to download with \texttt{curl} and the HTTP interface on the caches.
\end{enumerate}

If CVMFS is available, then it should be used first as it has the most redundant features, including: built-in GeoIP locating, rate monitoring, and fallback in failures.  XRootD client is used as well because it has efficient multi-threaded, multi-stream transfers of data from caches.  The fallback is using \texttt{curl}, which has the least features of the clients available.

In contrast to CVMFS, it does not provide a POSIX interface to the StashCache data federation.  \texttt{stashcp} cannot watch the application's read behavior to know what parts of files are required.  Instead, \texttt{stashcp} will copy the entire file from the cache to the execute machine.

Both the caches and origins are packaged by the Open Science Grid.  They are provided in RPM packages\cite{rpmcite}, Docker containers\cite{dockercontainer}, and Kubernetes\cite{kubernetes} configurations.

Further, we have responded to feedback from experienced StashCache administrators and created automatic configuration generation using a registration service.  An administrator simply specifies the experiments that the opportunistic storage should support and the configuration for a cache and origin is generated on-demand.

StashCache caches are installed at six universities and three Internet2 backbones.  In addition, a StashCache cache is installed at the University of Amsterdam.

\subsection{Monitoring}
We use XRootD's detailed monitoring in order to monitor and account for usage of StashCache.  Each StashCache cache sends a UDP packet for each file open, user login, and file close.  The collector of this information is complex since each packet contains different information.

\begin{itemize}
    \item \textbf{User Login:} The user login information includes the client hostname, the method of logging in, such as HTTP or xrootd protocol.  It also includes whether it was logged in with IPv6 or IPv4.  The user is later identified by a unique user ID number.
    \item \textbf{File Open:} Contains the file name, total file size, and the user ID which opened the file.  The file is later referred to by a unique file ID number.
    \item \textbf{File Close:} Contains the total bytes read or written to the file, as well as the number of IO operations performed on the file.  It contains the file ID from the file open event.
\end{itemize}

\begin{figure}[h]
	\centering
	\includegraphics[width=0.6\linewidth]{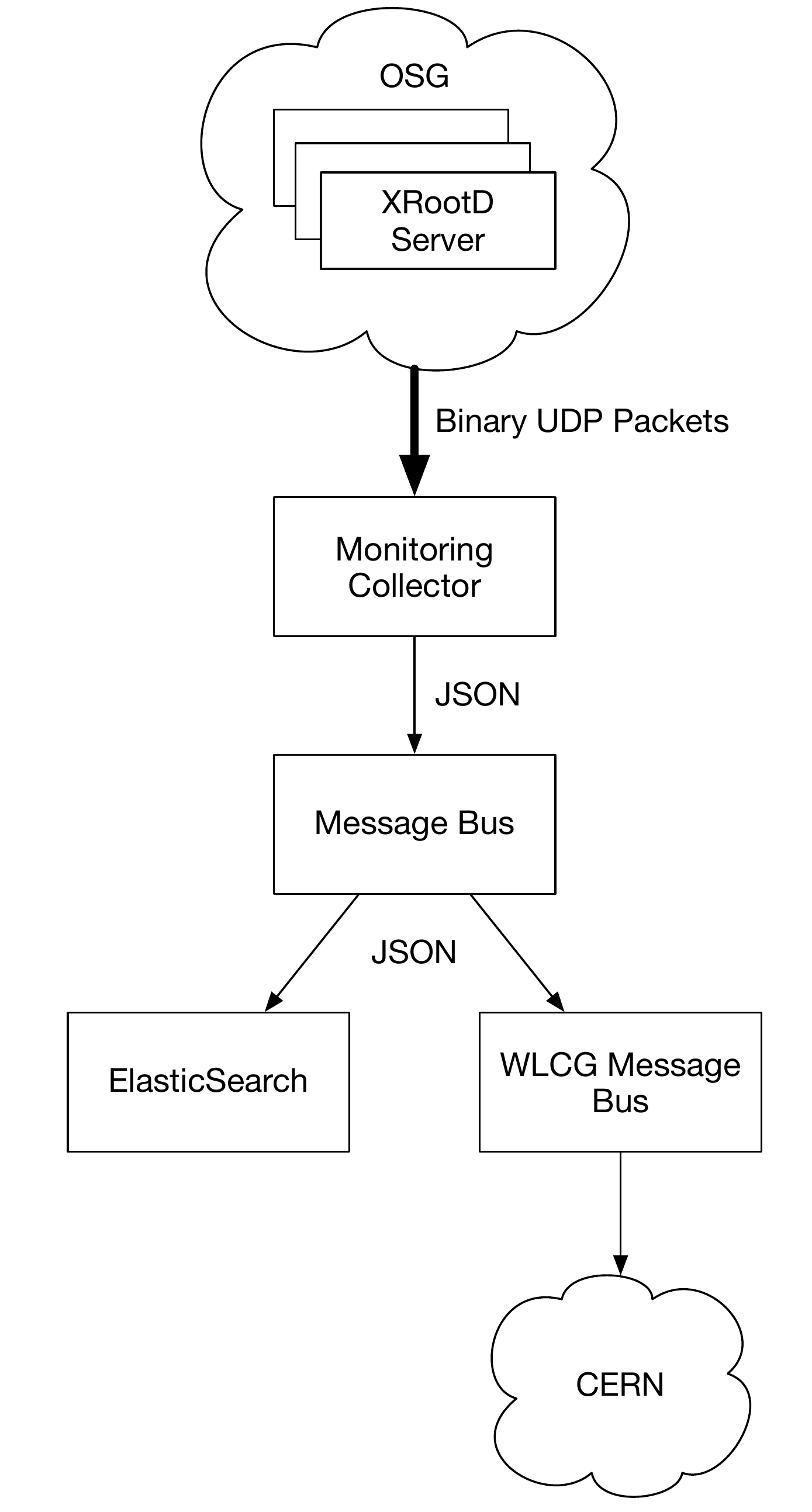}
	\caption{StashCache monitoring flow}
	\label{fig:cachemonitoringflow}
\end{figure}

Figure \ref{fig:cachemonitoringflow} shows the flow of monitoring information.  Each XRootD server sends binary UDP packets to the central monitoring collector.  The collector combines the different UDP packets to fill in full information for each file transfer.  On each file close packet, the collector combines the data from the file open and user login packets and sends a JSON message to the OSG message bus.

The OSG message bus distributes the file monitoring to databases in the OSG and the Worldwide LHC Computing Grid (WLCG).  The message is stored in the database for aggregation and analytics.

\begin{figure*}[h]
	\centering
	\includegraphics[width=0.8\textwidth]{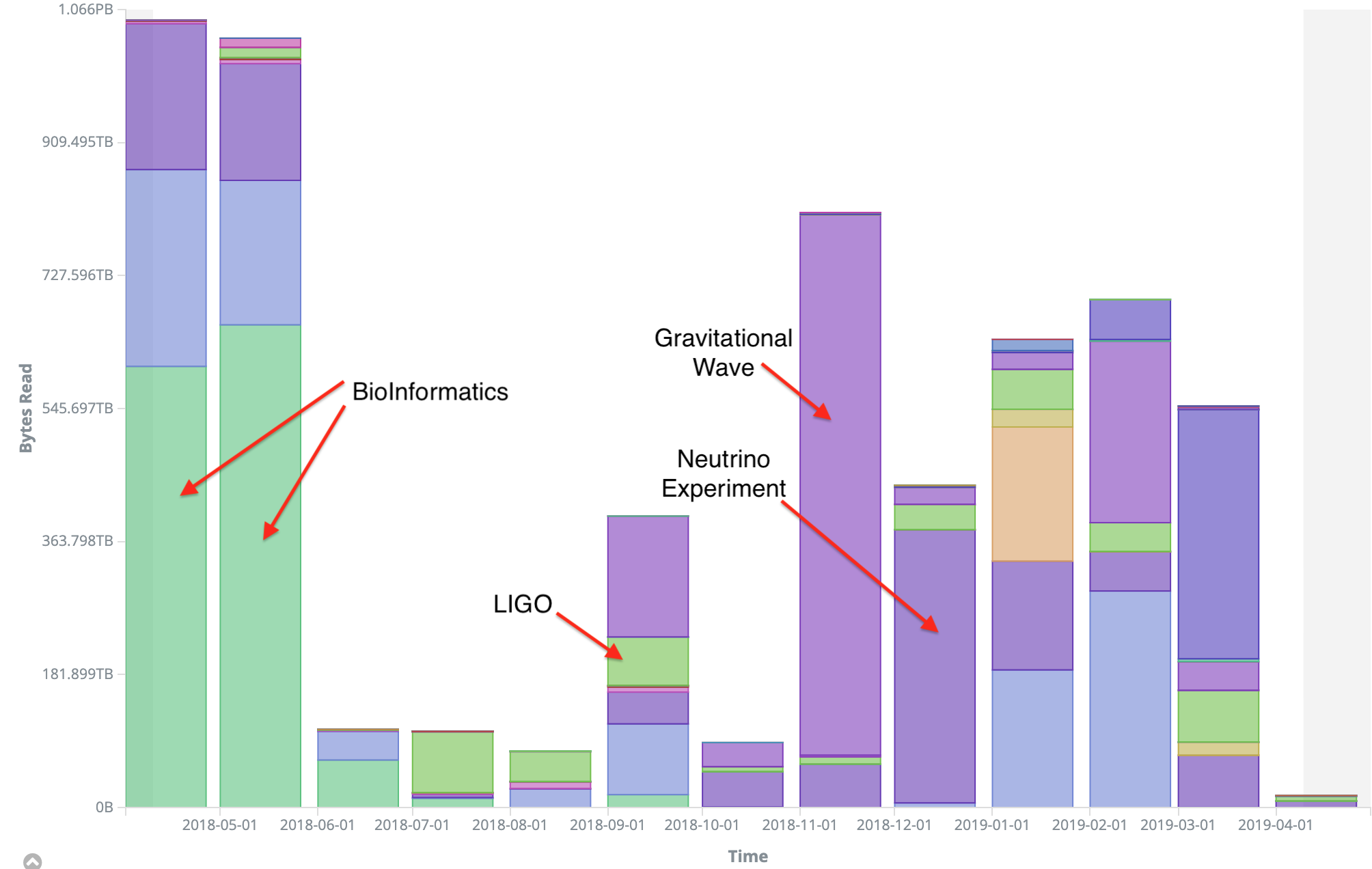}
	\caption{Last 1 Year of StashCache usage.}
	\label{fig:stashcacheutilization}
\end{figure*}

Figure \ref{fig:stashcacheutilization} shows the utilization of StashCache over the last year, from April 2018 to April 2019.

\section{Evaluation}

The benefits using StashCache are perceived by both the researchers and the resource owners.

Caches are located at sites that have volunteered opportunistic storage and on the Internet2 network backbone as part of the National Research Platform initiative.  Some sites have volunteered storage in order to help opportunistic users.  One site, Syracuse University, noticed wide area network usage to other caches and wanted to install a cache locally in order to minimize the network usage of outbound data requests.  After installation of a StashCache cache, the site noticed reduced wide area network connections.

\begin{figure}[ht]
    \centering
    \includegraphics[width=\linewidth]{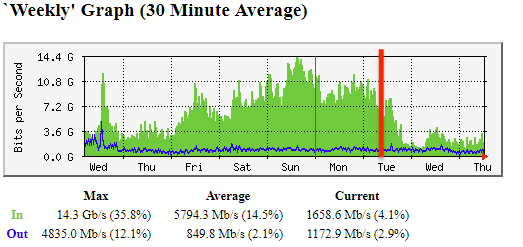}
    \caption{Syracuse Network Usage}
    \label{fig:syracuse-nework}
\end{figure}

Figure \ref{fig:syracuse-nework} shows the Syracuse WAN network bandwidth before and after StashCache was installed.  The bold red line shows when the StashCache server was installed.  Without the StashCache, Syracuse was downloading 14.3 GB/s of data.  After StashCache was installed, the network bandwidth reduced to 1.6 GB/s.  Not all of the data being monitored in this graph is StashCache data, other research data was captured; but the reduction in WAN network transfers is clear.

Many researchers use StashCache simply because there is no other way to transfer data to distributed computing resources.  They do not own storage resources that could scale to several thousand requests.  And they do not have relationships with sites to allow them to borrow storage.

Table \ref{tab:stashcacheusage} shows the top users of StashCache for 6 months ending in February 2019.


\rowcolors{2}{gray!25}{white}
\begin{table}[ht]
    \centering
    \begin{tabular}{l|l}
        \rowcolor{gray!50}
        \textbf{Experiment} & \textbf{Usage} \\ \hline
        Open Gravitational Wave Research & 1.079PB \\
        Dark Energy Survey & 709.051TB \\
        MINERvA (Neurtrino Experiment) & 514.794TB \\
        LIGO & 228.324TB \\
        Continuous Testing & 184.773TB \\
        NOvA & 24.317TB \\
        LSST & 18.966TB \\
        Bioinformatics & 17.566TB \\
        DUNE (Neutrino Experiment) & 11.677TB
        
    \end{tabular}
    \caption{StashCache Usage}
    \label{tab:stashcacheusage}
\end{table}

\subsection{Comparing StashCache to Distributed HTTP Proxies}

We tested the performance of StashCache and of HTTP proxies distributed at sites.  We theorize that HTTP proxies will have better performance with small files, while StashCache will be more efficient at large files.

For a test dataset, we created files with sizes representing the file size percentiles from the monitoring.  We queried the monitoring for the percentiles from the last 6 months of usage from October 2018 to April 2019.  The percentiles are shown in Table \ref{tab:stashcachepercentiles}.  Since the 95th and 99th percentile are the same value, we did not create a test file for the 99th percentile.  In addition to the percentiles, we also tested with a 10GB file which will show future potential for larger files.

\rowcolors{2}{gray!25}{white}
\begin{table}[ht]
    \centering
    \begin{tabular}{l|l}
        \rowcolor{gray!50}
        \textbf{Percentile} & \textbf{Filesize} \\ \hline
        1 & 5.797KB \\
        5 & 22.801MB \\
        25 & 170.131MB \\ 
        50 & 467.852MB \\
        75 & 493.337MB \\
        95 & 2.335GB \\
        99 & 2.335GB
    \end{tabular}
    \caption{StashCache filesize percentiles}
    \label{tab:stashcachepercentiles}
\end{table}

The test dataset was hosted on the Stash filesystem at the University of Chicago.  The files are available by HTTP and StashCache downloads while on the same filesystem.  There are many users of the filesystem, network, and data transfer nodes during our tests which provided realistic infrastructure conditions.

We chose to run against the top 5 sites providing opportunistic computing in the last six months on the OSG.  Those sites are Syracuse University, University of Colorado, Bellarmine University, University of Nebraska -- Lincoln, and the University of Chicago.

We created an HTCondor DAGMan workflow to submit the jobs to each site, without two sites running at the same time.  We avoided running two sites at the same time so there was no competition at the data origin for the files.

Each job downloads all files four times.  The first time it uses curl to download through the HTTP cache.  Since this is the first time downloading this unique file, it is assumed and verified that the first time is a cache miss.  It then downloads the file again through the HTTP proxy which will be a cache hit.

The third download is through \texttt{stashcp} and the StashCache federation.  This will download the file through a cache.  The fourth download is again using \texttt{stashcp}, but it should be cached.

\section{Results \& Discussion}

During the tests, there were many observations.  While running the experiments we experienced expiration of files within the HTTP proxies.  Our initial design of the experiments would loop through the list of download files, then loop again to download from the HTTP proxy.  After downloading the last large file, the first files were already expired within the cache and deleted.

The HTTP proxies at sites are configured to not cache large files.  In all of our tests, the 95th percentile file and the 10GB file were never cached by the HTTP proxies.  But they were cached by the StashCache caches.

\begin{figure}[ht]
    \centering
    \includegraphics[width=\linewidth]{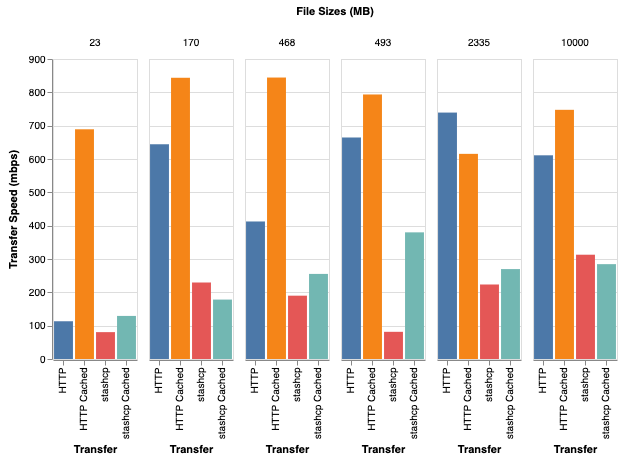}
    \caption{Colorado Cache Performance.  Higher is better.  Using the HTTP Proxies provide faster download speeds than using StashCache in all filesizes.  This could be because the HTTP proxy has fast networking to the wide area network, while the worker nodes have slower networking to the nearest StashCache cache.}
    \label{fig:colorado-performance}
\end{figure}

The tests made clear that some sites have very fast networking from the wide area network from to worker nodes, while others did not.  For example, see Figure \ref{fig:colorado-performance}.  The bandwidth between the worker nodes and the HTTP proxy is very good.  But, the bandwidth to the StashCache caches is consistent but lower than HTTP proxy performance.  Some sites prioritize bandwidth to the HTTP proxy, and between the proxy and worker nodes.  For example, they have larger bandwidth available from the wide area network to the HTTP proxy than to the worker nodes.

\begin{figure}[ht]
    \centering
    \includegraphics[width=\linewidth]{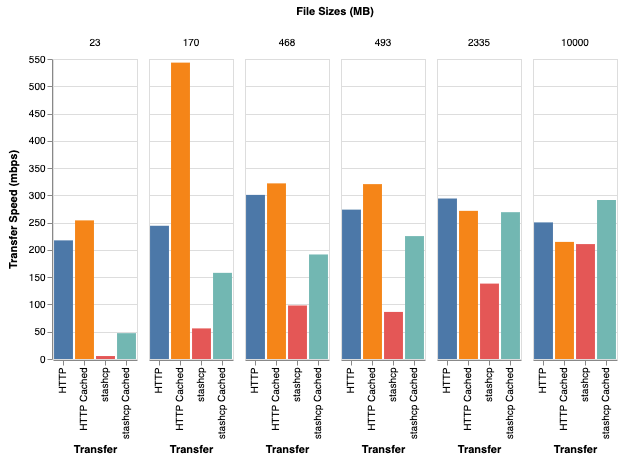}
    \caption{Syracuse Cache Performance.  Higher is better.  StashCache provides faster downloads for large files, but not for smaller files.}
    \label{fig:syracuse-performance}
\end{figure}

Compare Colorado's performance to Syracuse's performance in Figure \ref{fig:syracuse-performance}.  You will notice that the cached StashCache is always better than the non-cached.  Also, for large data transfers, StashCache is faster than HTTP proxies.  This indicates that the worker nodes have higher or equal bandwidth to the StashCache cache and the HTTP proxy.

\begin{figure}[ht]
    \centering
    \includegraphics[width=0.25\linewidth]{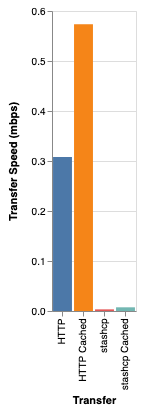}
    \caption{Small File Performance.  Higher is better.  This is downloading a 5.7KB file.  For this small of a file, HTTP performance is much better than StashCache.}
    \label{fig:small-file-performance}
\end{figure}

From the tests, it is clear that HTTP proxies provide superior transfer speeds for small files. In Figure \ref{fig:small-file-performance} you can notice that StashCache transfer speed for small files is always slower than the HTTP proxy.  \texttt{stashcp} has a larger startup time which decreases it's average performance.  The \texttt{stashcp} has to determine the nearest cache, which requires querying a remote server, then can start the transfer.  In contrast, the HTTP client has the nearest proxy provided to it from the environment.

For most of the tests, the very large file was downloaded faster with StashCache than a HTTP Proxy.  In table \ref{tab:httpvsstashcache}, you can see the decrease in time it takes to download larger files with StashCache vs. HTTP Proxy.  In the table, a negative value indicates a decrease in time to download the file through StashCache.  As mentioned above, Colorado stands alone with very fast performance for downloading through the HTTP proxy.

\rowcolors{2}{gray!25}{white}
\begin{table}[ht]
    \centering
    \begin{tabular}{l|l|l}
        \rowcolor{gray!50}
        \textbf{Site} & \textbf{2.3GB} & \textbf{10GB} \\ \hline
        Bellarmine & -68.5\% & -10.0\% \\
        Syracuse & +0.9\% & -26.3\% \\
        Colorado & +506.5\% & +245.9\% \\
        Nebraska & -12.1\% & -2.1\% \\
        Chicago & +30.6\% & -7.7\% \\
    \end{tabular}
    \caption{StashCache HTTP Proxies vs StashCache.  This is the percent difference between the HTTP Proxy and StashCache.  Negative values indicate the time to download decreased when using StashCache}
    \label{tab:httpvsstashcache}
\end{table}

The tools and analysis notebooks are available \cite{derek_weitzel_2019_2635588}.

\section{Conclusions and Future Work}
StashCache provides a scalable and transparent method for data distribution for opportunistic users of cyberinfrastructure.  From our analysis, we have shown that StashCache has better performance than HTTP proxies for large files.  But for small files less than 500MB, HTTP proxies provide better performance.

StashCache provides features not available when using HTTP proxies, such as a posix interface with CVMFS.  Also, CVMFS calculates checksums of the data, which guarantees consistency of the data which HTTP proxies do not provide.  Also, the HTTP proxy cache can expire rapidly which can cause re-downloads of the data from the origin.

Future testing will be needed to confirm our findings in a variety of infrastructure conditions.  We ran our tests over the course of several days and the results may be different in the future.  We have no visibility into the resource contention of the network, caches, proxies, or origin server.  Future tests should be run over a longer period of time which will show the performance of StashCache over many conditions.

In the future, we hope to add more capabilities to the StashCache federation.  Specifically, we plan to support output data in a write-back cache configuration.  Writeback cache will allow users to write output files to a cache rather than back to the origin.  Once the files are written to StashCache, writing to the origin will be scheduled in order to not overwhelm the origin.





%

%
\begin{acks}
This research was done using resources provided by the Open Science Grid \cite{pordes2007open}, which is supported by the National Science Foundation award 1148698, and the U.S. Department of Energy's Office of Science. This work was supported by the National Science Foundation under Cooperative Agreement OAC-1836650.
\end{acks}
%
\bibliographystyle{ACM-Reference-Format}
\bibliography{StashCache.bib}

\end{document}